\documentclass[12pt]{article}
\usepackage{graphicx}
\newcommand{\drp}[2]{\frac{\partial{#1}}{\partial{#2}}}

\title{\Large {\bf The  Schr\"{o}dinger  equation, the
    zero-point electromagnetic radiation and the photoelectric effect.}}

\author{ H. M. Fran\c{c}a$^1$\footnote{e-mail: hfranca@if.usp.br},\,
  A.  Kamimura$^2$ and G. A. Barreto$^2$ \\{\footnotesize $^1$
    Instituto de F\'{i}sica, Universidade de S\~{a}o Paulo C.P.  66318
    , 05315-970 S\~{a}o Paulo, SP,
    Brazil}\\
  {\footnotesize$^2$ Instituto de Eletrot\'{e}cnica e Energia,
    Universidade de S\~{a}o Paulo,} \\{\footnotesize Av.  Prof.
    Luciano Gualberto, 1289 CEP 055080-010, S\~{a}o Paulo,SP, Brazil}}

\begin{document}
\maketitle
{\bf Abstract}

A Schr\"{o}dinger type equation for a mathematical probability
amplitude $\Psi(x,t)$, is derived from the generalized phase space
Liouville equation valid for the motion of a microscopic particle,
with mass $M$, moving in a potential $V(x)$. The  particle phase space probability density is denoted $W(x,p,t)$ and the entire system is immersed in the ``vacuum'' zero-point electromagnetic radiation . We show that the 
generalized Liouville equation is reduced to a non-quantized
Liouville equation in the equilibrium limit where the small radiative
corrections cancel each other approximately. Our
derivation will be based on a simple Fourier transform of the non-quantized phase space probability distribution $W(x,p,t)$. For
convenience, we introduce in this Fourier transform an auxiliary constant $\alpha$, with dimension
of action, and an auxiliary coordinate denoted by $y$. We shall prove that $\alpha$ is equal to the Planck's constant present in the momentum operator of the Schr\"{o}dinger equation. Moreover, we shall show that this momentum operator is  deeply related with the ubiquitous zero-point electromagnetic
 radiation. It is also important to say that we do not assume that the mathematical amplitude
$\Psi(x,t)$ is a de Broglie matter-wave, in other words, the  wave-particle duality
hypothesis is not used within our work.  The implications of our study
for the standard interpretation of the photoelectric effect is
discussed by considering the main characteristics of the phenomenon. We also mention, briefly, the effects of the zero-point radiation in the tunneling phenomenon and the Compton's effect.

{\it{} Key words}: Foundations of quantum mechanics; Schr\"{o}dinger
 equation; Zero-point electromagnetic fluctuations; Photoelectric
effect.


\newpage

\section*{1. Introduction}
\bigskip

In 1932 E. Wigner published an important paper {\cite{wigner,moyal}} where he
introduced what is called today {\it{} the Wigner's quasi probability function}, or simply
{\it{} Wigner's phase space function}. We shall denote it by $Q(x,p,t)$ and, according to the Wigner proposal,
it is connected with a Schr\"{o}dinger  equation solution $\Psi(x,t)$ by the expression:

\begin{equation}\label{wf}
Q(x,p,t) \equiv \frac{1}{\pi\hbar}\int _{- \infty}^{+ \infty} \Psi ^*(x-y,t)
\Psi(x+y,t) e^{-\frac{2ipy}{\hbar}}dy,
\end{equation}
where $\hbar$ is the Planck's constant. Wigner's intention was to
obtain a quantum mechanical description of the phase space microscopic
and macroscopic phenomena as demanded by the {\it{}correspondence
  principle} \cite{manfredi,hayakawa,francab}.  We recall that the
Schr\"{o}dinger equation, namely

\begin{equation}\label{schuras}
{i\hbar\drp{\ }{t}\Psi(x,t) = \left[ \frac{-\hbar^{2}}{2M}\drp{^{2}}{x^{2}} + V(x)\right]\Psi(x,t)\quad,}
\end{equation}
also depends explicitly on the Planck's constant $\hbar$ \cite{sch}.

We want to emphazise that neither Wigner nor Schr\"{o}dinger clarified
the relation of the constant $\hbar$ with the classical stochastic
zero-point electromagnetic radiation. Notice, however, that the
existence of the classical stochastic zero-point radiation was 
considered {\it seriously} by Albert Einstein and Otto Stern in the
period 1911-1913 \cite{otto}, and by  T. W. Marshall in an important
paper published in 1963 \cite{roy}. Interesting experimental observations of the ``vacuum'' zero-point fluctuations are described in more recent publications \cite{amir}.

We also would like to mention that, if applied to an arbitrary
solution $\Psi(x,t)$ of the Schr\"{o}dinger equation (\ref{schuras}),
the expression (\ref{wf}) may lead to {\it{} negative} phase space
probability densities. In other words, the relation (\ref{wf}) should
be considered only as the definition of an {\it{}auxiliary} \cite{hmf_twm}
function, which has some properties of a classical distribution,
useful to calculate the statistics on the configuration and momentum
space. Interesting examples are the excited states of the harmonic
oscillator \cite{hmf_twm}.

 The definition (\ref{wf}), and the Schr\"{o}dinger equation
(\ref{schuras}), do not lead to the classical Liouville equation
except when $V(x)$ is quadratic in the variable $x$ (see ref.
\cite{manfredi}). However, according to the correspondence principle,
one should necessarily obtain the Liouville equation in {\it{}the case
  of a macroscopic particle} moving in a {\it generic} potential $V(x)$.

We recall that our method was suggested by some authors in the past
\cite{hayakawa,pena2}. Here, in contrast with the original Wigner
procedure, we start from a {\it non-quantized Liouville}
equation in order to obtain a {\it  Schr\"{o}dinger type
  equation} \cite{pena,ori,genesis}.

 In other words, starting with the {\it generalized Liouville
  equation in phase space}, we shall obtain the mathematical and the
physical conditions necessary to relate it with an equation that is
formally identical to the Schr\"{o}dinger equation for a {\it{}
  mathematical probability amplitude} $\Psi(x,t)$ appropriately defined
in the configuration space.

 With the use of the Wigner transform, that is, the inverse Fourier
transform of the  equation (\ref{wf}),  we shall disclose the
{\it{}conditions} in which the relation between an approximate Liouville equation
and the Schr\"{o}dinger type equation is valid for a {\it generic} potential $V(x)$. 

 Our presentation is organized as follows. We give, within section 2,
the physical background and the mathematical approximations necessary
to connect the non-quantized phase space probability distribution to the
mathematical configuration space probability amplitude $\Psi(x,t)$. The
relation between the corresponding Liouville equation and the  Schr\"{o}dinger
 equation is explained in details within the section 3. The
role of the random zero-point radiation with spectral distribution
$\rho_0(\omega)$ given by \cite{pena2,pena}

\begin{equation}\label{r0}
\rho _0 (\omega)\equiv \lim _{T \rightarrow 0} \left[
  \frac{\omega^2}{\pi^2c^3}\left( \frac{\hbar \omega}{2}+ \frac{\hbar
      \omega}{\exp{\left( \frac{\hbar
            \omega}{kT}\right)-1}}\right)\right]= \hbar \omega^3/2\pi^2c^3,
\end{equation}
is also explained within the section 3. Some implications of our study
for  the standard  interpretation of the photoelectric
  effect are presented in the section 4 . The section 5 is devoted to a
brief discussion in which we mention the role of $\rho_0 (\omega)$
in the free particle propagator, in the tunneling phenomenon and in
the Compton effect.


\section*{2. Physical background and mathematical approximations}
\label{sec:two}
\bigskip

Our starting point is the non-relativistic Liouville equation for the
classical probability distribution in phase space, denoted by $W(x,p,t)$. This
probability density is the {\it average} \cite{pena} distribution over the
realizations of the {\it fluctuating electric field} indicated by $E_x(t)$
in the equation (\ref{Md}) below. The generalized Liouville equation,
associated with the non-relativistic motion of a particle with mass
$M$ and charge $e$, evolves in time according to \cite{pena,ori,genesis}
\begin{eqnarray} \label{hayakawa}
\frac{\partial}{\partial t} W(x,p,t)+\frac{p}{M} \frac{\partial\;
  W(x,p,t)}{\partial x}+\frac{\partial}{\partial p}\left[ F(x) +
  \frac{\gamma}{M}F^{\prime}(x)p\right] W(x,p,t)\simeq \nonumber \\ \simeq e^2
D(t)\frac{\partial ^2}{\partial p^2} W(x,p,t)\;\;.
\end{eqnarray}

Here $F(x)$ is the deterministic force, $\gamma$ is a very short time interval ($\gamma=
\frac{2}{3}\frac{e^2}{Mc^3}\simeq 6.3 \times 10^{-24}\;$ sec$\;$ for
electrons) and $e^2D(t)$ is the time dependent diffusion coefficient. According to de la Pe\~{n}a and Cetto \cite{pena} the
eq. (\ref{hayakawa}) is a non-Markoffian Fokker-Planck equation with
memory (see ref. \cite{pena} for the details). We recall that the term
proportional to $\gamma$ in (\ref{hayakawa}) is due to the radiation reaction force.

 The equation (\ref{hayakawa}) was derived by assuming that
the total force acting on the charged particle is given by

\begin{equation}\label{Md}
 M \frac{d^2 x}{d t^2}= F(x) + e E_x(t) + \frac{2}{3}\frac{e^2}{c^3} \stackrel{\dots}{x}(t)+ ... \;\; ,
\end{equation}
where $E_x(t)$ is the ``vacuum''electric field in the dipole
approximation. This random field is such that $\left < E_x (t)\right >
= 0$ on average. The reader can see the work of T. H. Boyer
\cite{boyerb} for an expression of $E_x(t)$ in terms of plane waves
with random phases. The last term in (\ref{Md}) is of order $e^2$, and
is the first approximation of the radiation reaction force.

 According to the analysis presented by
L. de la Pe\~{n}a and A. M. Cetto (see the sections II,
III and V of their paper \cite{pena} ) it is possible to conclude that
(\ref{hayakawa}) is equivalent to the non-quantized Liouville equation 

\begin{equation}\label{pw}
\frac{\partial W(x,p,t)}{\partial t} + \frac{p}{M}\frac{\partial W(x,p,t)}{\partial
   x}+ F(x) \frac{\partial W(x,p,t)}{\partial p} \simeq 0 \;\; ,
\end{equation}
in the equilibrium limit where the radiation reaction term and the diffusion term in
(\ref{hayakawa}) {\it compensate} each other in an approximate way \cite{pena}.

 After the above clarifications we shall use the simpler
equation (\ref{pw}) in  order to relate it with the
Schr\"{o}dinger type equation for the mathematical probability
amplitude $\Psi(x,t)$. The approximate equation (\ref{pw}) is much more easy to
handle than the generalized Liouville equation (\ref{hayakawa}).

 Following S. Hayakawa \cite{hayakawa} we shall consider the
following Fourier transform defined by
\begin{equation}\label{wigner-ft}
{\widetilde{W}(x,y,t) \equiv \int_{-\infty}^{\infty}{dpW(x,p,t)e^{\frac{2ipy}{\alpha}}},}
\end{equation}
where y is the auxiliary coordinate,  $\alpha$ is the auxiliary constant
with dimension of action, and $W(x,p,t)$ is the  phase space
distribution function (see the equation (\ref{pw})). The constant $\alpha$ will be
easily fixed later with the use of  physical and mathematical criteria.

 We know that $W(x,p,t)$ is always real and positive definite and the classical configuration space probability density can be defined mathematically as 

\begin{equation}\label{pdex1}
P(x,t) \equiv \int dp W(x,p,t)= \lim _{y\rightarrow 0}\widetilde{W}(x,y,t).
\end{equation}

 Notice that this probability density is the {\it average}
distribution over the realizations of the {\it fluctuating} electric field
indicated by $E_x(t)$ in the equation (\ref{Md}).

 Since $P(x,t)$ is  real and positive it is {\it always
possible} to introduce a mathematical probability amplitude $\Psi(x,t)$
as \cite{pena,ori,genesis}

\begin{equation}\label{pdex2}
P(x,t)\equiv \left | \Psi(x,t)\right |^2 = \Psi^{*}(x,t) \Psi(x,t) \; ,
\end{equation}
where the factorization $ \Psi^{*}(x,t) \Psi(x,t)$ is completely
general in the $limit \;y \rightarrow 0$. In other words, the
equations (\ref{pdex1}) and (\ref{pdex2}) can be summarized as
\begin{equation}\label{psi2}
  \Psi^*(x,t)\Psi(x,t)\equiv\lim  _{y\rightarrow
    0}\int^{\infty}_{-\infty}dpW(x,p,t)e^{\frac{2ipy}{\alpha}}
  \equiv\lim _{y\rightarrow 0} \widetilde{W}(x,y,t), \end{equation}
where the limiting process considered above means that the behavior
of $\widetilde{W}(x,y,t)$ when $y \approx 0$ is {\it very important}.
However, the above definition (\ref{psi2}) is not enough to determine
a differential equation for the amplitude $\Psi(x,t)$. In order to
achieve this goal we shall use the approximate Liouville equation
(\ref{pw}). We shall see, in the next section, that the approximate Liouville
equation   has a {\it close relation} with the
Schr\"{o}dinger type equation for the mathematical probability amplitude $\Psi(x,t)$.


\section*{3. The approximate Liouville equation and its connection with the
Schr\"{o}dinger  equation}\label{sec:three}

It is a widespread belief \cite{manfredi} that the connection between
the Wigner quasi probability function (\ref{wf}) , the Liouville
equation (\ref{pw}), and the Schr\"{o}dinger equation (\ref{schuras}), is only
possible  for quadratic potentials in the $x$ variable.{\it We
shall extend this connection to a more general potential} $V(x)$,{ \it that
is, beyond the case of the harmonic potential}. In order to achieve
this goal we shall use a procedure similar to that presented  by
S. Hayakawa \cite{hayakawa} and  K. Dechoum, H. M. Fran\c{c}a and C. P. Malta
\cite{francab}. Dechoum, Fran\c{c}a and Malta discussed several classical aspects of the
{\it Pauli-Schr\"{o}dinger equation} written in the spinorial notation. Here
we are treating the spinless case.

To obtain the differential equation for $\Psi(x,t)$ we shall use the
Wigner type transformation defined previously in the equation
(\ref{wigner-ft}). Regarded as a simple mathematical definition, we
stress that this Fourier transform contains  the same dynamical
information carried by the phase space density $W(x,p,t)$.

 We observe that, due to the definitions (\ref{wigner-ft}), (\ref{pdex1}), (\ref{pdex2})
and (\ref{psi2}), the Wigner type transform $\widetilde{W}(x,y,t)$ is
a complex function {\it that has a physical significance only in the limit} $|y|\rightarrow 0$, that is,

\begin{equation}\label{qx}
\lim_{y\rightarrow 0} \widetilde{W}(x,y,t)= \left|\Psi(x,t) \right|^2=
\Psi(x,t)\Psi^*(x,t).
\end{equation}

 For this reason, we shall consider, in what follows,  the
definition (\ref{wigner-ft}) only for {\it small values} of $y$ (see
also the reference \cite{edel} for a  similar approach). Our goal is to obtain the differential equation for $\Psi(x,t)$, from the differential equation for $\widetilde{W}(x,y,t)$, in the limit of small $y$.

 Our first step is to consider the equation
\begin{equation}\label{dwtiudt1}
\frac{\partial}{\partial t}\widetilde{W}(x,y,t)=\int^{\infty}_{-\infty} \frac{\partial W}{\partial t}
e^{\frac{2ipy}{\alpha}}dp .
\end{equation}

 \noindent Using the Liouville equation (\ref{pw}) and  after an integration by
 parts, we obtain
\begin{equation}\label{dwtiudt2}
       \begin{array}{lr}
\frac{\partial \widetilde{W}}{\partial t}=-\int^{\infty}_{-\infty}
\left[
\frac{p}{M}\frac{\partial W}{\partial x}+F(x)\frac{\partial W}{\partial p}
\right]
e^{\frac{2ipy}{\alpha}}dp = &\\ \\ = \left [ -\frac{i\alpha}{2M}\frac{\partial^2}{\partial_y\partial_x}+F(x)\frac{2iy}{\alpha}\right ]\int^{\infty}_{-\infty} W(x,p,t)e^{\frac{2ipy}{\alpha}}dp.
      \end{array}
\end{equation}

 Substituting (\ref{wigner-ft}) into (\ref{dwtiudt2}) we get
 the Hayakawa \cite{hayakawa} equation for $\widetilde{W}(x,y,t)$, namely

\begin{equation}\label{dwtiudt4}
i\alpha\frac{\partial}{\partial t}\widetilde{W}(x,y,t)=
\left[
\frac{(-i\alpha)^2}{2M}\frac{\partial ^2}{\partial y\partial x}
-2yF(x)
\right]\widetilde{W}(x,y,t) .
\end{equation}

In order to facilitate the calculations it is convenient to use new
variables, namely $s=x-y$ and $r=x+y$, so that the equation
(\ref{dwtiudt4}) can be written as

\begin{equation}\label{dpsipsidt1}
	i\alpha \frac{\partial}{\partial t}\widetilde{W}(r,s,t)= \left[\frac{(-i\alpha) ^2}{2M}
\left(\frac{\partial ^2}{\partial r^2}-\frac{\partial ^2}{\partial s^2}\right)
-(r-s)F
\left(
\frac{r+s}{2}
\right)
\right]\widetilde{W}(r,s,t) .
\end{equation}

 According to (\ref{qx}), we want an equation for
$\widetilde{W}(r,s,t)$, when $y = (r -s)/2\to 0$, that is, when $r\to
s$. Consequently we must consider that the points $r$ and $s$ are
{\it{}arbitrarily} close in the equation  (\ref{dpsipsidt1}). Therefore, in accordance to the {\it{} mean value theorem}, one can consider that

\begin{equation}\label{tvm}
-(r-s)F
\left(
\frac{r+s}{2}
\right)
\simeq -\int _{s}^{r}F(\xi)d\xi \simeq V(r)-V(s) ,
\end{equation}
is a very good approximation \cite{hayakawa}. Notice that the mean value theorem requires only that the integrand function (force $F(x)=-V'(x)$ in this case) is a continuous  function in the small integration interval under consideration. 

 The use of equation (\ref{tvm})  will imply in a  great
simplification of the problem of finding the differential equation for
the mathematical probability amplitude $\Psi(x,t)$, because
(\ref{dpsipsidt1}) becomes {\it separable} in the variables $r$ and $s$ as we
shall see below. Substituting (\ref{tvm}) into (\ref{dpsipsidt1}) we get the equation

\begin{equation}\label{schrod1}
i \alpha \frac{\partial}{\partial t}\widetilde{W}(r,s,t) = \left[\frac{(-i\alpha)^2}{2M}\left(\frac{\partial}{\partial r^{2}} - \frac{\partial}{\partial s^{2}}\right) + V(r) - V(s)\right]\widetilde{W}(r,s,t)\quad,
\end{equation}
which was obtained by S. Hayakawa  in 1965 (see also
ref. \cite{pena}). One can verify that the above equation is {\it{}separable}
in the variables $r$ and $s$ if we introduce a function
$\widetilde{W}(r,s,t)$ such that

\begin{equation}\label{schrod2}
\widetilde{W}(r,s,t) \equiv \Psi^{*}(s,t)\Psi(r,t).
\end{equation}
Notice that (\ref{schrod2}) is in accordance with our previous
equation (\ref{pdex2}). Thus (\ref{schrod1}) is decomposed into the equation

\begin{equation}\label{schrod4}
i \alpha \frac{\partial}{\partial t}\Psi(x,t)=
\left[
\frac{1}{2M}
\left(
-i \alpha \frac{\partial}{\partial x}
\right)^2 + V(x)
\right]\Psi(x,t) ,
\end{equation}
and its complex conjugate. The equation (\ref{schrod4}) is formally
identical to the Schr\"{o}dinger equation provided that the operator
$-i\alpha\partial/\partial x$ is identified with the Schr\"{o}dinger
momentum operator $p=-i\hbar\partial/\partial x$ , that is,  the
  auxiliary constant $\alpha$ is identified with  $\hbar$. See 
  the  paragraphs containing our equations (\ref{nonreleqmvto}),(\ref{comutador1}) and (\ref{momento}), where we discuss this point using the Heisenberg picture. 

Using a more explicit notation we get, from the equation (\ref{schrod4}), the result
\begin{equation}\label{ih}
i \hbar \frac{\partial \Psi}{\partial t} (x,t)= \left[ \frac{1}{2M} \left( -i\hbar
    \frac{\partial}{\partial x}\right)^2 + V(x)\right] \Psi(x,t)\equiv
\hat{H} \Psi(x,t),
\end{equation}
which has exactly the form of the  Schr\"{o}dinger equation for {\it probability amplitude} $\Psi(x,t)$. Notice that
$\hat{H}$ is the Hamiltonian operator of the system and the equation  (\ref{ih}) possesses a
{\it complete set} of {\it solutions},
$\Psi_n(x,t) \equiv \phi_n(x) \exp(-i\epsilon_n t/\hbar)$, where the functions $\phi_n(x)$ are such that 
\begin{equation}\label{ham}
 \hat{H} \phi_n(x) \equiv \epsilon_n \phi_n(x) \; 
\end{equation}
The constants $\epsilon_n$ are the ``energy levels'' and the functions $\phi_n(x)$ are the ``eigenfunctions'' of the system \cite{schiff}.

We stress that, due to our approximations, the
excited states of the set $\{\epsilon_n,\phi_n(x)\}$ {\it do not
  decay}  (see the references
\cite{hyd} and \cite{dalibard} for the physical 
explanation of the decay processes). It is important to recall that the equation (\ref{ih}) was derived from the non-quantized Liouville equation (\ref{pw}) for the phase space probability distribution $W(x,p,t)$.


 We know that the Planck's constant $\hbar$ is related with the
{\it{} electromagnetic fluctuation phenomena} characteristic of
 Stochastic
Electrodynamics (SED) \cite{pena,boyerb} and Quantum Electrodynamics (QED) \cite{dalibard}. The most important of these
fluctuations are associated with the zero temperature electromagnetic
radiation which has a spectral distribution such that

\begin{equation}\label{rozero}
\rho _0(\omega)=\frac{\hbar\omega^3}{2\pi^2c^3} \;,
\end{equation}
in both QED and SED. We recall that these two theories have
many features in common \cite{dalibard,boyerb}. It is also interesting to
recall that the {\it classical} effects of the zero-point radiation
were discovered very early by M. Planck in the period 1911-1912 \cite{pena2}. The quantum zero-point radiation was discovered  latter, in
1927, by Paul Dirac.

Some properties of the spectral distribution $\rho_{0}(\omega)$ will
be explicitly used at this point. For the moment we would like to say
that, using the {\it{}Heisenberg picture}, Sokolov and Tumanov
\cite{sokolov} and P.W Milonni \cite{milonni1,milonni2} were able to
show that the commutation relation between the position operator
$x(t)$ and the momentum operator $p(t)$ are strongly related to $\rho
_0(\omega)$ given in the equation (\ref{rozero}). Moreover, according
to de la Pe\~{n}a, Vald\'{e}s-Hernandez and Cetto \cite{val}, who use
a {\it non-quantized} system as a starting point, the action of the
electric zero-point field on matter is essential and ultimately leads
the matrix (or Heisenberg) formulation of quantum mechanics for the
motion in an {\it arbitrary} potential $V(x)$. This is an important
conclusion. Therefore, according to these authors, a ``free''\,\,
electron (mass $M$ and charge $e$) has a nonrelativistic equation of
motion for the position operator $x(t)$ such that

\begin{equation}\label{nonreleqmvto}
{M\ddot{x}(t) = \frac{2}{3}\frac{e^{2}}{c^3}\stackrel{\dots}{x}(t) + eE_{x}(t),}
\end{equation}
within the  Heisenberg picture \cite{sokolov,milonni1,milonni2,val}.
Recall that, according to the authors quoted in ref. \cite{val},
$E_{x}(t)$ can be a { \it fluctuating non-quantized electric field}.
We shall use the  Heisenberg approach in what follows.

 Within the  Heisenberg
picture the equation (\ref{nonreleqmvto}) can be easily solved.
We get the position operator $x(t)$ and the canonical momentum
operator $p(t)\equiv M\dot{x} + \frac{e}{c}A_{x}(t)$, where $A_{x}(t)$
is the quantized vector potential
$\left(E_{x}=-\frac{1}{c}\drp{A_{x}}{t}\right)$.

 From the solution for $x(t)$ we get $p(t)$. Therefore, it is
possible to calculate the commutation relation between the position
operator $x(t)$ and the canonical momentum operator $p(t)$. The result
is

\begin{equation}\label{comutador1}
\left[\; x(t),p(t)\;\right] = \left[x,M\dot{x}\right] = 4 i \pi^2 c^3 \gamma \int _0^{\infty} d\omega
\frac{\rho _0(\omega)}{\omega^3(1 + \gamma ^2 \omega^2 )} = i
\hbar\;,
\end{equation}
where $\gamma = 2 e^2 / 3 M c^3$. This is an interesting and relevant
result because only $\rho_{0}(\omega)$ depends on $\hbar$ in the above
integral. Moreover, the result (\ref{comutador1}) is independent
of the charge of the particle. Since the Heisenberg and the
Schr\"{o}dinger pictures are equivalent (see the references
\cite{critical} and \cite{vacuum} for a good discussion of this point) one concludes that the momentum
operator used in the Schr\"{o}dinger type equation (\ref{ih}), namely
\begin{equation}\label{momento}
p = -i \hbar \partial/\partial x \,\,, 
\end{equation}
is {\it deeply related with the ``vacuum'' electromagnetic fluctuations}
with spectral distribution $\rho _0 = \hbar \omega^3/2 \pi^2 c^3$. 


\section*{4. The  photoelectric effect and the zero-point radiation}
\bigskip

A three-dimensional generalization of the one-dimensional  Schr\"{o}dinger type
equation (\ref{ih}) is

\begin{equation}\label{3d}
i\hbar \frac{\partial \Psi(\vec{r},t)}{\partial t}= \left[ \frac{-\hbar^
  2}{2M} \nabla^2 - V(\vec{r}) -  e \vec{r}\cdot \vec{E}_c (\vec{r},t)\right]\Psi(\vec{r},t),
\end{equation}
where we have used the {\it dipole} approximation. The term $ -  e
\vec{r}\cdot \vec{E}_c (\vec{r},t)  $ is the interaction energy, $e$
is the electron charge and $M$ is the electron mass. 

In the above equation $\vec{E}_c(\vec{r},t)$ is a {\it classical deterministic}
electric field, associated with the {\it incident} radiation, and
$V(\vec{r})$ is the atomic Coulomb potential for instance. We are not
explicitly considering the stochastic ``vacuum'' fields in the
equation (\ref{3d}) because it was concluded, in previous works
\cite{critical,vacuum}, that this may generate double counting.{ \it  The operator $-i
\hbar \vec {\nabla}$  already contains the effects of
the zero-point electromagnetic fields necessary to our purposes}. 

An interesting physical feature of the above three-dimensional
generalization of the mathematical Schr\"{o}dinger type equation
(\ref{ih}) is that it leads to a novel   interpretation of
the photoelectric effect.

 The reader can find a good discussion of this
point in the subsection entitled ``{\it The photoelectric effect}''
(pgs. 40, 41 and 42) of the article by M. O. Scully and M. Sargent III
published in 1972 \cite{scully}. Whithin this paper it is used a
{\it{}classical deterministic} electric field, of a monochromatic plane wave,
polarized in the $z$ direction, which is denoted $\vec{E}_c(\vec{r},t)
= \hat{z} E_0 \cos(\omega t - k y)$, where $E_0$ is the amplitude of
the electric field and $\omega$ is the corresponding angular frequency
($\omega = kc$).
 
According to an standard analysis, based on the equation (\ref{3d})
and on the ``Fermi Golden Rule'', these authors obtained the expression
\cite{scully,lamb}

\begin{equation}
\frac{d P_f}{dt} = 2 \pi \left|\left<f\left| e\vec{r} \cdot \hat{z} \right|g\right>\right|^{2} \hbar^2 E_0^2 \;t\;
\delta \left( \omega - \frac{\epsilon _f - \epsilon _g}{\hbar} \right)\;,
\label{rate}
\end{equation}
for the probabilistic rate of emission of photo-electrons  from a bounded
initial state of energy $\epsilon _g$ to the final state of energy
$\epsilon _f = Mv^2/2 $, where $\epsilon_f$ is the continuous energy
of the ejected electron, and $e\vec{r}$ is the electric
dipole of the atom. Notice, however, that the authors mentioned in references \cite{scully} and \cite{lamb} do not mention the role of the zero-point radiation.

From the argument of the delta function in
(\ref{rate}), we see that

\begin{equation}
\omega = \frac{\epsilon _f - \epsilon _g}{\hbar}\;.
\label{freq}
\end{equation}
The Planck's constant $\hbar$ appearing above was introduced within the
mathematical  Schr\"{o}dinger type equation (\ref{3d}), without  using the
concept of  {\it photon}. Moreover, according to the
interpretation of the Schr\"{o}dinger equation presented within our
paper (see the section 3 ) the origin of $\hbar$ can be traced back to
the expressions (\ref{comutador1}) and (\ref{momento}) which\, involves
$\rho_0(\omega)$, that is, {\it the Planck's constant, in the equation
  (\ref{freq}), has its \, origin in the {\it zero-point radiation}
  with spectral \, distribution} $\,\,\rho_0 = \hbar
\omega^{3}/(2\pi^{2} c^{3})$. Therefore, the famous Einstein equation for the
photoelectric phenomenon follows from (\ref{freq}) and can be put in the
form \cite{scully,lamb}.

\begin{equation}\label{fotoefeito}
\hbar \omega =  \epsilon _f - \epsilon _g \geq  \frac{Mv^2}{2} + \phi  \;,
\end{equation}
where $\phi$ is the work function  characteristic of the
material, and $Mv^2/2$ is the continuous kinetic energy of the ejected electron. See our figure 1 for an illustration .
\begin{figure}[!h]
\begin{center}
\includegraphics[scale=0.50]{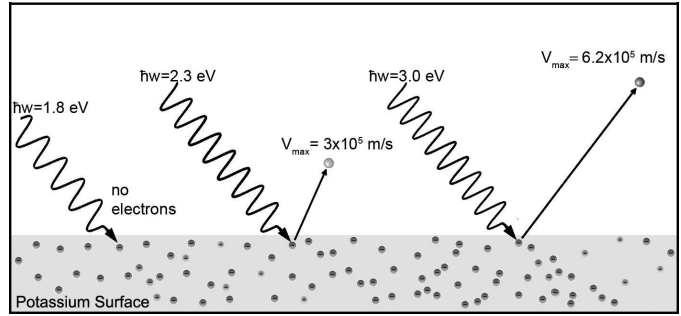}
\caption{\small { \it \it Illustration of the photoelectric effect
    from a potassium surface} ($2.0 \; eV${\it needed to eject
    electrons}). {\it In the figure we indicate the characteristic
    energy $\hbar \omega$ of the ``photons''  . For instance $\hbar
    \omega = 1.8 \; eV$, $\hbar \omega = 2.3 \; eV$ and $\hbar \omega
    = 3.1 \; eV$. The maximum velocity $\;V_{max}$ of the ejected
    electrons,is indicated for each frequency $\omega$} ({\it see the
    Einstein equation (\ref{fotoefeito}})). Notice that the electrons inside the metal move with different energies.}
\end{center}
\end{figure}

In order to better describe
the ``collision''(or scattering) process we need the equations for the {\it momentum
  conservation}. We recall that Einstein introduced  the {\it classical concept} of
``needle radiation'' \cite{pena2} which has ``momentum and energy''.  See our figure 2 for an illustration.
\begin{figure}[!h]
\begin{center}
\includegraphics[scale=0.52]{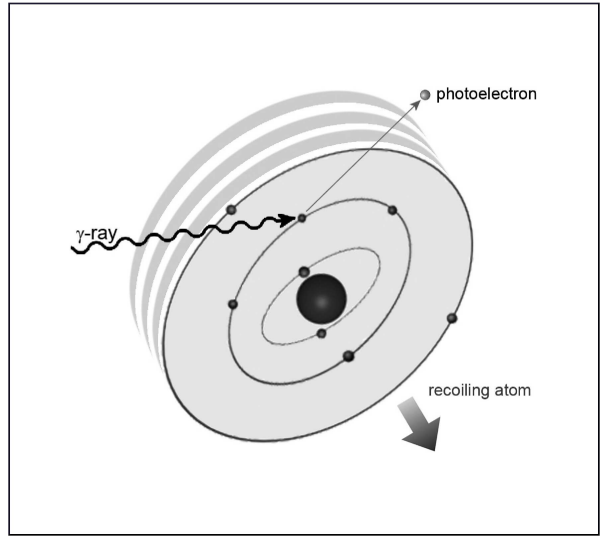}
\caption{\small \it Illustration of the momentum conservation
  in the photoelectric phenomenon. In this case the atom will suffer a
  recoil in the indicated direction. In other words, the parent atom
  (or crystal) plays an important role in the photoelectric effect (see
  the  section III of the reference \cite{kidd}).}
\end{center}
\end{figure}

The role of the {\it momentum conservation equations} is very well
illustrated in the section III of the paper by R. Kidd, J. Ardini and
A. Anton \cite{kidd}. These authors gave very interesting examples
relating the {\it velocity variation} of the recoiling electron, and
the {\it direction of the recoil} of the parent atom. In other words,
it is shown that for some values of $\hbar \omega$ and $\phi$ the value
of the momentum of the recoiling electron can be {\it much larger}
than the momentum of the incident ``photon'' due to the recoil of the
parent atom (see the figure 2 in the section III of the reference \cite{kidd}).

Consequently, the Einstein's equation for the photoelectric effect can be
interpreted in a  non-quantized manner  where the zero-point
electromagnetic field play  a very important role.  This opens a
new way of understanding the photoelectric effect. It is our intention
to explore this interpretation in future calculations, by considering
the atoms in {\it special physical conditions} (flying between very
close to conducting Casimir plates \cite{fms}, localized close to some
element of an electric circuit \cite{bdfs} or moving inside a
solid material \cite{blanconew}). In these cases there are important
modifications in the spectral distribution $\rho_0(\omega)$.

\section*{5. Brief discussion  }

We want to recall that within this paper we have clarified the deep
relationship between the Schr\"{o}dinger equation, the zero-point
electromagnetic radiation and the photoelectric effect. Another
important observation is that the kinetic energy operator present in
the unidimensional equation (\ref{ih}), namely

\begin{equation}\label{kinetic}
\frac{1}{2M}\left(-i \hbar\frac{\partial}{\partial x} \right)^2 ,
\end{equation}

\noindent generates the  spreading of the mathematical probability amplitude
$\Psi(x,t)$. In other words, if $V(x) = 0$ we get from the
equation (\ref{ih}) the  ``free'' particle  propagator \cite{eugen}

\begin{equation}\label{propagador1}
K(x,t|x_0,t_0)=\sqrt{\frac{M}{2\pi i\hbar (t-t_0)}}\exp\left[\frac{iM(x-x_0)^2}{2\hbar (t-t_0)} \right]\,,
\end{equation}
 
\noindent which connects $ \Psi(x_0,t_0)$ with $ \Psi(x,t)$ in the time
interval $t-t_{0}$, that is

\begin{equation}\label{psi20}
\Psi(x,t)= \int_{-\infty}^{\infty}dx_0 \Psi(x_0,t_0) K(x,t|x_0,t_0).
\end{equation}
This solution was obtained from the mathematical Schr\"{o}dinger 
equation (\ref{ih}) assuming that $V(x)=0$. The above equations mean that, even in
the far ``empty'' space, for example, the matter particles are always
interacting with the vacuum zero-point radiation. The
inevitable conclusion is that the fluctuating electromagnetic
background with spectral density $\rho_{0}(w) = \hbar w^{3}/2\pi
c^{3}$ is the source of the {\it zero temperature  diffusion}. In
other words, the constant $\hbar$ present in the propagator
$K(x,t|x_0,t_0)$ has its origin in the spectral density
$\rho_0(\omega)$. Notice that the equations (\ref{propagador1}), (\ref{psi20}) and the {\it superposition principle} suggest that the interference phenomena of eletrons, and other matter particles, may be {\it explained whithout the use of de Broglie waves}.

We also would like to say that, from the connection between the
approximate Liouville equation (\ref{pw}) and the Schr\"{o}dinger type
equation (\ref{ih}) presented here, one can conclude that the
mathematical probability density $\left |\Psi(x,t)\right |^2$ also may be
not necessarily associated with real waves \cite{alfred}. Moreover,
according to the {\it Copenhagen Interpretation}, the amplitude
$\Psi(x,t)$ is not a physical wave but a ``probability wave''
\cite{pais}. As a matter of fact, the existence of de Broglie waves
was recently questioned by Sulcs, Gilbert and Osborne \cite{sulcs} in
their analysis of the experiments, by M. Arndt et al. \cite{arndt}, on
the interference of {\it{}massive} particles as the $C_{60}$ molecules
(fullerenes). According to M. Arndt et al. \cite{arndt} ``the de
Broglie wavelength of the interfering fullerenes is already smaller
than the diameter of a single molecule by a factor of almost 400''. This is an important observation which
deserves further attention.

It is equally important to mention that, recently, a { \it tunneling}
phenomenon was also interpreted in  classical terms, that is, without
using {\it the concepts of de Broglie waves} and the {\it wave-particle duality
hypothesis} \cite{tunneling}. The important concept used was, again,
the existence of the zero-point radiation with spectral distribution
given by $\rho_0(\omega)=\hbar w^3/2\pi^2c^3$.

Finally we want to say a few words concerning the relationship between
the Compton effect and the zero-point radiation. This relationship was
established  by H. M. Fran\c{c}a and A. V. Barranco \cite{tein} in 1992,
by studying a system of free molecules in equilibrium with thermal and
zero-point radiation. 

In their work Barranco and Fran\c{c}a, using a statistical approach, replaced the Einstein concept of
{\it random spontaneous emission} by the concept of  {\it stimulated emission
by the random zero-point electromagnetic fields} with spectral distribution $\rho_0(\omega)$. As a result, Compton and Debye's
kinematic relations were obtained within the realm of a completely
classical theory, that is, without having to consider the
wave-particle duality hypothesis for the molecules or the
radiation bath. This is another important dynamical effect of the
zero-point radiation.We recall that many other relevant effects of the
zero-point electromagnetic fields are presented in the excellent book
by L. de la Pe\~{n}a and A. M. Cetto  \cite{pena2}.

\bigskip
\newpage
{\bf Acknowledgement}

\noindent 
We thank Prof. Coraci P. Malta, Prof. L. de la Pe\~{n}a,
Prof. A. M. Cetto and Prof. Geraldo F. Burani  for
valuable comments and J.S.Borges for the help in the preparation of
the manuscript. We also acknowledge Funda\c{c}\~{a}o de Amparo à
Pesquisa do Estado de S\~{a}o Paulo (FAPESP) and Conselho Nacional de
Desenvolvimento Cient\'{i}fico e Tecnol\'{o}gico (CNPq) for partial
financial support.

\end{document}